# An Empirical Exploration of ChatGPT's Ability to Support Problem Formulation Tasks for Mission Engineering and a Documentation of its Performance Variability

**Max Ofsa[a], Taylan G. Topcu[b]**
[a]Grado Department of Industrial and Systems Engineering, Virginia Tech, maxo@vt.edu
[b]Grado Department of Industrial and Systems Engineering, Virginia Tech, ttopcu@vt.edu

**Abstract**

***Systems engineering (SE) is evolving with the availability of generative artificial intelligence (AI) and the demand for a systems-of-systems perspective, formalized under the purview of mission engineering (ME) in the US Department of Defense. Formulating ME problems is challenging because they are open-ended exercises that involve translation of ill-defined problems into well-defined ones that are amenable for engineering development. It remains to be seen to which extent AI could assist problem formulation objectives. To that end, this paper explores the quality and consistency of multi-purpose Large Language Models (LLM) in supporting ME problem formulation tasks, specifically focusing on stakeholder identification. We identify a relevant reference problem, a NASA space mission design challenge, and document ChatGPT-3.5's ability to perform stakeholder identification tasks. We execute multiple parallel attempts and qualitatively evaluate LLM outputs, focusing on both their quality and variability. Our findings portray a nuanced picture. We find that the LLM performs well in identifying human-focused stakeholders but poorly in recognizing external systems and environmental factors, despite explicit efforts to account for these. Additionally, LLMs struggle with preserving the desired level of abstraction and exhibit a tendency to produce solution specific outputs that are inappropriate for problem formulation. More importantly, we document great variability among parallel threads, highlighting that LLM outputs should be used with caution, ideally by adopting a stochastic view of their abilities. Overall, our findings suggest that, while ChatGPT could reduce some expert workload, its lack of consistency and domain understanding may limit its reliability for problem formulation tasks.***

## 1. Introduction

Increasing demand towards a more interconnected world pushes the Systems Engineering (SE) community [1,2,3] towards a Systems-of-Systems (SoS) perspective [4]. A recent policy change in the US Department of Defense (DoD) formalizes these trends around ME [5], where the goal is to architect a heterogenous portfolio of SoS with independent constituents that may dynamically evolve in terms of their roles and integration level depending on the nature of the mission [6]. This view considerably increases the complexity of the problem space due to its combinatorial nature [7, 8]. However, as documented by numerous US Government Accountability Office Reports, SE community has notoriously struggled with dealing with local problems that are constrained within the boundary of a standalone system [9, 11]. Methods and tools for dealing with this increasing ME complexity, albeit a few exceptions [12, 13], is nascent.

Among the new methods and technologies that are helping to push the envelope of how ME can be accomplished effectively are various forms of Artificial Intelligence (AI). From generative design tools [14, 15], to AI assistants [16], and semantic databases [17, 18], these rigorous tools are providing previously unobtainable capabilities by collaborating with their users to synthesize vast amounts of data into accessible forms where they might be used to preform trade space analysis or iterate through hundreds of design variations rapidly. More recently, LLMs like the Generative Pre-trained Transformers (GPTs), such as ChatGPT, have heated the conversation around the capabilities and usefulness of AI advancements [19]. From passing the bar exam to making meal plans, the flexibility and effectiveness of LLMs raise expectations about their potential use for almost any activity [20, 21]. ME is no exception to this new wave.

However, a fundamental question remains. If confronted with tasks that require knowledge-based performance for open-ended questions, can a multi-purpose LLM like Chat GPT provide adequate outputs [22]? One of the core challenges of SE and ME is problem formulation, that revolves around translating ill-defined problems into well-defined ones for which an engineering solution can be developed [23, 24]. This process is inherently shaped by domain knowledge regarding the nature of how complex systems must be derived [25, 26, 27]. The point of departure here is that an AI agent, particularly one that is general purpose, might not be able to provide reliable answers, but instead, might be more lenient towards gathering data and presenting it in an intelligible manner [28]. In short, the efficacy of multi-purpose AI tools to in terms of providing “skilled” enough feedback [21, 29] similar to one that would be expected by a human SE expert is nascent [30].

In this study, we empirically evaluate the ability of open access LLMs to assist with ME tasks, more specifically the fundamental problem formulation task of stakeholder identification. We choose to focus on problem formulation due to two factors. First, it is essential to framing any SE or ME problem [23, 31, 32] and their goodness often dictate the efficacy of the rest of the development process [2, 33]. The second reason is their potential value in terms of freeing up domain-experts [34]. To elaborate, problem formulation tasks like this are often handled by a team of domain-experts [35, 36] that have significant tacit knowledge regarding the context [37] and are fluent in the “art” of systems engineering [38]. However, these experts are highly valued and become costly to utilize partly due to their limited numbers [39]. Shifting some of the domain-expert’s workload [40] to AI could potentially free-up valuable expert time, speed up design processes, and enable effective utilization of more novice engineers [34, 41]. Thus, we chose a representative space mission problem provided by NASA’s RASC-AL program [42] and probe some LLMs ability to effectively serve this objective.

This paper pursues two research questions. First, for the reasons identified above, can multi-purpose LLMs help with ME problem formulation tasks such as stakeholder identification? Second, given that LLMs are not deterministic tools, how reliable and consistent is an LLM for a given stakeholder identification task, and what is the typical variance in its responses? Due to the black-box nature of LLMs, we contend that our second research question is particularly important as our ability to trust LLM responses depend on their consistency, as much as their accuracy. Large variance across independent iterations may indicate that every information acquisition attempt is subject to uncertainty; thus, one may need a stochastic approach before utilizing LLM outputs.

Our findings reveal nuanced results. We find that AI assistance in problem formulation tasks may be useful to domain experts; through leveraging novice level knowledge about the broader context and presenting it in palatable form that could lessen expert workload. These outputs come at a cost of being muddled by variance in both the volume of usable and unusable results. Encouragingly, our small data set found very few instances of purely wrong responses. Nevertheless, we also find that the depth of knowledge of an open-access LLM like Chat GPT may be somewhat ill-suited for these open-ended tasks.

## 2. Literature Review

### 2.1. Mission Engineering

Historically, systems were acquisitioned as standalone entities (e.g., a tank, a missile, a communication satellite) and were later integrated in teams to execute strategic missions (e.g., an integrated task force). In late 2010s, the US DoD published the Mission Engineering Guide [43], to outline an SoS engineering approach that prioritizes mission success in an attempt to mitigate the lack of mission understanding that leads to interoperability and integration issues, and uncertainties in fielded capabilities. Thus, ME, at its core, is a form of SoS engineering [43]. The primary difference between ME and standard SoS practices is inclusion of operational context at all levels of development. ME is more specifically defined as: *"Planning, analysing, organizing, and integrating current and emerging operational concepts for the purpose of evolving the end-to-end operational architecture and capability attributes, across the Doctrine, Organization, Training, Materiel, Leadership, Personnel, and Facilities (DOTMLPF) spectrum, including anticipated Blue Force (BLUFOR) and Opposition Force (OPFOR) behaviours, that are needed to inform the communities of interest involved in fulfilling mission needs statements* [44]."

In addition to the growing awareness in the DoD, there is an increasing demand for ME in other domains. For instance, while DoD's policy guidance is rather new, NASA has been working on space missions with a similar mission oriented view [45]; and has its own robust mission guidance document [1]. Similarly, the private industry is in the processes of commercializing the Low Earth Orbit environment, the Federal Emergency Management Agency (FEMA) outlines its policy goals around five overarching missions [46], and domains covered by the Cybersecurity and Infrastructure Security Agency (CISA) are some of the many places where ME is in demand [47].

### 2.2. Translating Ill-defined Problems into Well-defined Ones, Its Importance for Mission Engineering

Every technical or programmatic solution starts with formulating it as a problem, and this applies to both SE and ME [43, 48]. Specifically, creating a well-defined problem [24, 27] involves identifying the boundary of the system of interest, along with its associated stakeholders, and their needs, preferences, and constraints [48]. Without rigorous execution of these steps, resulting system requirements suffer from either unproper bounding, underdevelopment, or overdesign, which leads to a solution that would be unable to meet mission needs [49, 50]. Particularly in the case of SoS, improper early stage bounding render downstream decomposition decisions ineffective [51, 53], regardless of the time and energy invested, and often results in significant rework [54, 55].

### 2.3. Human-AI Collaboration for Systems Engineering

Wide-spread AI use is already under heavy scrutiny from nearly every discipline. Within SE we can find examples like Selva and Martin's work in building an open source virtual assistant to allow ease of work on the designer [16]. Sarica et al. have explored the use of large semantic networks of technical terms for patent information, like TechNet, to drive the generation of new ideas base on those that already exist within the database [17]. Chen et.al. are engaging with generative adversarial networks and how they can be tuned to synthesize high-quality and diverse designs, while allowing exploration of a portion of the design space during concept development [56]. Chong et al. delves into the confidence and cognitive abilities of AI-assisted humans as they make decisions, finding that poor AI recommendations lead to diminished confidence in the use of AI in decision making [57]. These studies support the notion that domain expertise plays an important role in the influence of AI recommendations on the goodness of a decision. Furthermore, Doris et. al. shows that, while multi-modal LLMs show promise, there is still a need for development in these models to reliably interpret complex engineering documents especially for tasks requiring precise retrieval and cross-referencing of information across various formats [58].

## 3. Methodology

We provide an overview of our research approach in Fig. 1. In step 1a we select an appropriate reference mission engineering problem from NASA RASC-AL challenges, in 1b, we choose a problem formulation task, which in this case is stakeholder identification. Step 2 iteratively refines prompts. In Step 2a, we create a task specific prompt and in 2b we create independent parallel threads and query an available LLM, which in this case was ChatGPT 3.5. In Step 2c, we briefly analyze the outputs, and either refine the prompt in Step2d and repeat Step 2, or if we deem the output reasonable, we pass the LLM output into Step 3, ex-post qualitative analysis. Below we elaborate each of the steps.

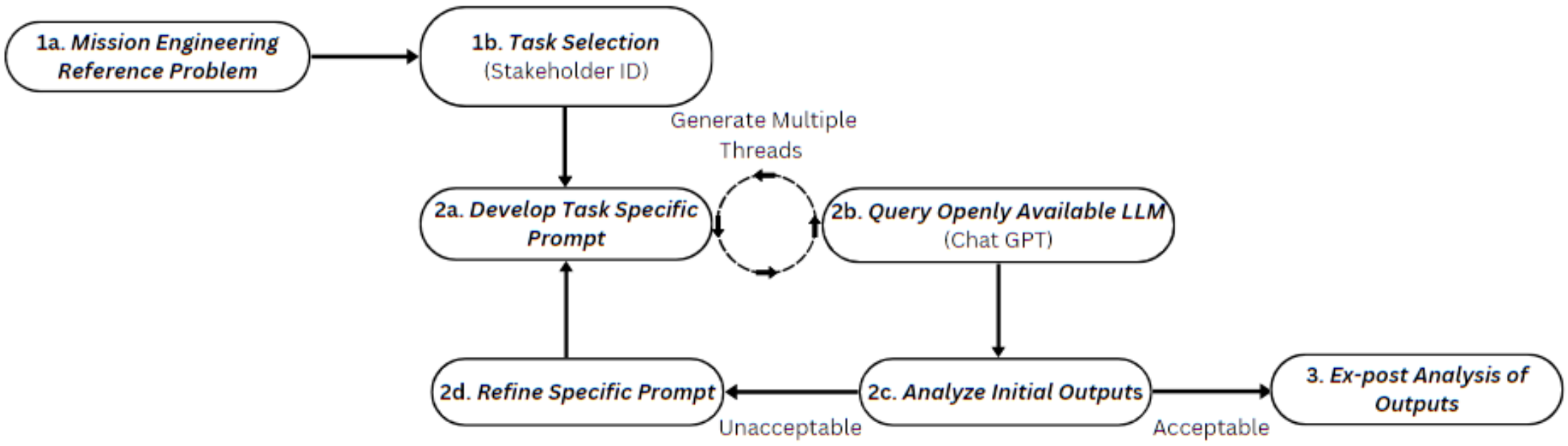

Fig. 1. Research Methodology

### 3.1. Step 1 – Selection of the Reference Problem and Mission Engineering Task

In Step 1a we chose a reference problem. For our interests NASA's RASC-AL competitions provided an ideal representative case as their problem statements refer to complex SoS problems with uncertain assets and resources. Furthermore, RASC-AL problems include well-defined mission need statements that have clear mission goals. We chose the 2021 Ceres reference mission, because at the time that this research was conducted, Chat GPT's database contained no information beyond September of 2021 [59]. The refence mission need statement reads:

*"Develop a concept that supports a crew of 4 on a mission to Ceres in the 2040s. The mission concept should take advantage of having crew in proximity to Ceres for low latency science operations, identify what planetary science payloads could be delivered during the mission, and include the ability for at least two crew to land on the surface of Ceres. Consider long duration health effects including but not limited to radiation, microgravity, and isolation. Mission should be < 5 years total, ready to land on Ceres no later than December 31, 2049, with an annual budget of no more than $3B/year from 2035 to proposed mission end date."* [42]

### 3.2. Step 2 – Initializing, Prompt Formulation, and Iterative Evaluation of Intermediate LLM Outputs

The relevancy of outputs from a given LLM can be skewed by lack of context, lack of framing, and/or too many or too few inputs [29, 60]. If a prompt lacks proper context or is not initialized to give the LLM a certain role, the resulting outputs can tend towards irrelevant content. Alternatively, if a prompt lacks proper framing, resulting outputs may be broad and unspecified. Thus, initializing and prompt development were critical for our purposes. We addressed this issue with an iterative approach.

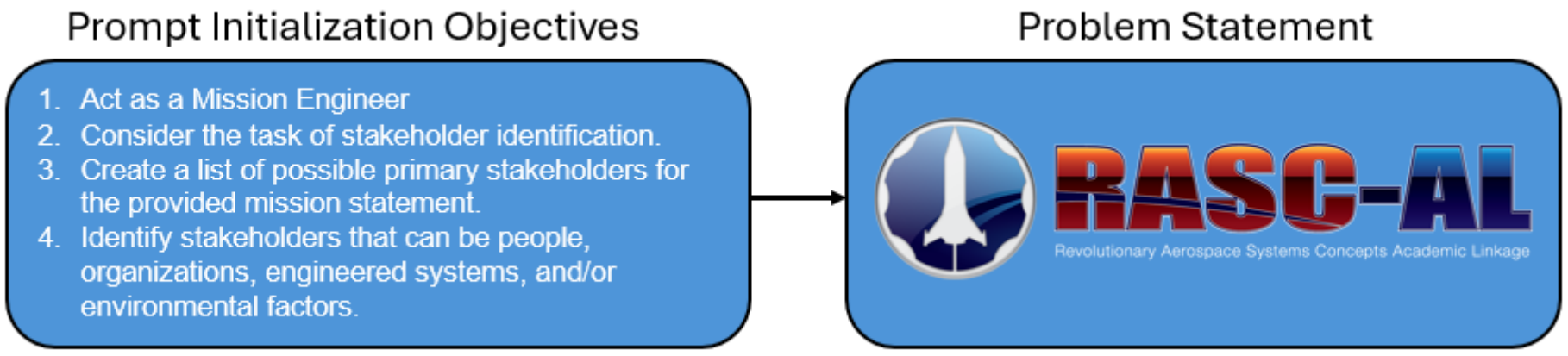

Fig 2 Prompt Initialization Objectives

In Step 2a we initialize the LLM as outlined by Fig. 2. and start our prompting. If a prompting thread contains previous prompts, we have observed that results can begin to suffer from having too much user-provided information to consider. In return, results generated through continued threads tend towards repeating themselves (or echoing) when asking new but potentially related questions [60]. Thus, in this study each prompt was generated in a new thread to avoid recursion of information in the outputs. Additionally, we used the "context manager" prompting pattern which initializes the LLM into the role of a SE by stating "Within the scope of Systems Engineering...". This initialization was followed by that request for the LLM to "...please consider the following:". Finally, the need statements provided by NASA RASC-AL contain enough specificity to remove the need for an additionally reframing; therefore, we only added prompt context to the needs statement and proceeded with providing these prompts to the LLM. In Step 2b we provide the prompts to an LLM of choice, which in this case was OpenAI's ChatGPT-3.5.

This selection was made based on research that identified it as highly capable of answering questions related to STEM associated work [61]. At the time of this work ChatGPT 4 was available via subscription so it was not considered as we were comparing freely available LLMs. It is also important to note that LLMs are not deterministic tools, and their outputs may include a ton of variability. Thus, to observe variance in outputs, for each of our prompts we created threads of ten parallel independent inquiries.

Expanding on Step 2b, we brought focus to problem formulation tasks within mission engineering by initializing accordingly. For this task, we specified a focus on stakeholders and the forms they might take, persons, organizations, engineered systems, or environments, in following with industry heuristics [2].

In Step 2c, we assessed the outputs from Step 2b for their alignment with our ground truth list of stakeholders. Prompts generating relevant outputs were advanced to Step 3 for further analysis, while less aligned prompts were refined in Step 2d to improve relevance. The goal was not to achieve perfect answers but to document the usefulness of LLM outputs with well-crafted prompts, focusing on relevance rather than optimization.

### 3.3. Step 3 – Ex-Post Analysis of Results

We analyzed each thread, by parsing its stakeholder information in a spreadsheet then qualitatively comparing each of the identified stakeholders for the following criteria:

- **Correctness:** through a direct comparison to the ground truth list of stakeholders.
- **Traceability:** a check that the provided stakeholder is real and appropriate for the needs statement.
- **Specificity:** level of abstraction of the identified stakeholder. Used to remove stakeholders that would not be part of initial problem formulation.

These metrics begin to validate a set of responses generated by Chat GPT and allows for a distilled view of each thread for comparison. As part of the correctness evaluation, Table 1 shows a ground truth list of 12 high level stakeholders was developed in response to the needs statement we chose. Due to the explorative nature of our research, these stakeholders were derived without distinction within the hierarchy of stakeholder types. Early attempts at prompting primary, secondary, and sponsor stakeholder types proved to be ineffective.

| **Stakeholder** | **Description** |
|---|---|
| Mission Crew | Persons who will operate mission vehicles and perform mission tasks |
| Mission Operations | Persons/groups/divisions that oversee the various phases of the mission lifecycle. Science, engineering, medical, and other units that develop and monitor mission systems. |
| Transportation and Launch Operations | Physical equipment and procedures associated with vehicle ground transportation and launch. (Launch platforms, trains, ships, staging/stacking facilities, etc.) |
| Collaborating Mission Assets | Organizational systems and/or existing equipment that may be leveraged for the mission. Such as crew or mission components from commercial entities. |
| Vehicle Component Manufacturers | Organizations that will manufacture and produce the needed vehicles for the mission. |
| Communications Networks | Infrastructure required to interface with to maintain communications throughout the mission like the Deep Space Network. |
| Operational Environment | The environments the vehicle and crew will need to survive to complete the mission. |
| NASA Administration | Identified as the oversight group that would handle budget distribution and develop the overall goals for the mission. |
| Collaborating Agencies and Governments | Any international collaborators that will have a stake in the pursuits and/or outcomes of the mission or provide crew and assets to the mission. |
| The United States Government | The body that sponsors the mission and provides oversight for continued funding. |
| Regulatory Committees | Organizations that dictate policy, safety, and well-being of persons and environments impacted by space missions. |
| General Public and Academia | Civilian groups or organizations that have a vested interest in the outcomes of the mission. |

*Table 1. Ground Truth Stakeholders List*

For each of the output threads, identified stakeholders were evaluated qualitatively using the three dimensions of correctness, traceability, and specificity. Although there are analytical methods to achieve this, such as cosine similarities, in this study we adopted a qualitative approach to avoid the discrepancy between the terms used to define a stakeholder vs. what the actual LLM definition conveyed. For example, the "Astronaut Corps" might be identified instead of "Mission Crew"; with very similar descriptions and justification. Here, to differentiate our identification, we used the traceability evaluation, which helped to separate correct responses from wrong. In the ground truth "Mission Crew" is defined as "the crew selected to perform the mission". However, in the "Astronaut Corps" response may be described as "…the committee that selects crew to perform the mission"; which in this case reads very similar but is semantically wrong.

On the other hand, the specificity metric differentiates the outputs into wrong or, more commonly, over-specified responses. For example, commonly a group like "NASA Human Research Program" would be identified with language like "Division responsible for addressing the long-duration health effects on the crew…." While this falls in with the "Mission Operations" stakeholder, it is a very specific aspect of mission operations. This program may indeed be considered a stakeholder within the mission lifecycle as objectives for the mission are defined, however at the initial problem formulation, this is not the case. Lastly, missed ground truth stakeholders were identified by simply subtracting the total number of correctly identified stakeholders from the total number of ground truth responses.

## 4. Results

### *4.1.* Output Comparisons

We complied the 25 LLM threads for visual comparison, marking stakeholder outputs in each as correct, over specified, wrong, and missed. Fig 3 is a stacked bar chart where each bar represents an individual thread from ChatGPT,s and colors represent stakeholders. Green stands for correct, over-specified in orange, wrong in pink, and red represents missed stakeholders compared to the ground truth.

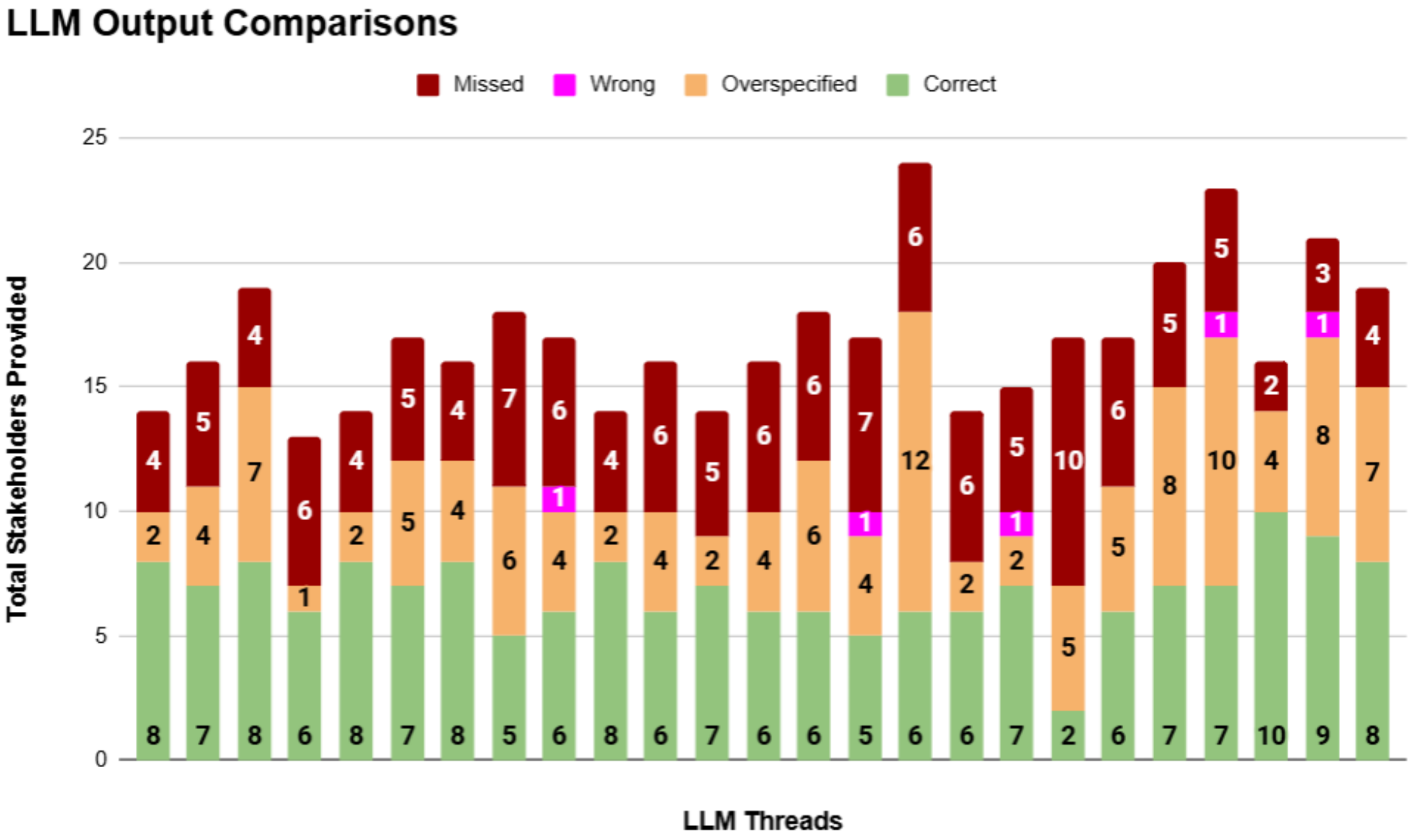

Fig. 3. LLM Output Comparisons

A review of the threads in Fig 3 suggest that ChatGPT never provided a completely correct or completely wrong output; as indicated by bars consisting of mixed colors. Nearly half of the outputs provided less outputs than the 12 ground truth stakeholders. On average, the LLM identified ~7 stakeholders correctly, though which specific stakeholders were captured in each thread varied. Interestingly, LLM was always able to identify two stakeholders correctly in each attempt: the NASA Administration and the Mission Operations stakeholder.

Overall, we found that GPT was consistently good at accurately identify stakeholders that are either people or organizations that could be “spoken to” during mission development. For instance, the LLM commonly, identified Vehicle Component Manufacturers, Collaborating Space Agencies and Governments, and the General Public and Academia stakeholders. This may be a result of stakeholder definitions that tend to favor human elements in their identification. However, the Mission Crew and Regulatory Committees were two human focused stakeholders that were identified roughly half the time. The Mission Crew was often missed; and commonly confused with a selection committee for astronauts, rather than the astronauts themselves.

In contrast, environmental factors and external systems were always left out of identified stakeholders; despite explicit prompting as: “…a stakeholder can be a human, organization, engineered system, or environmental factor…”. For instance, the “Operational Environment” dictates the operational conditions and influences requirements similar to any other stakeholder; however, was not identified by any of the threads. Similarly, environmental conditions on Earth play a significant role in both the deployment of physical assets as well as the ability to launch; yet never identified. We contend that this might be a result of the common tendency among practitioners to disregard “non-human” elements from the list of stakeholders. If such documents were used in LLMs’ training corpus, it might be disregarding non-human stakeholders. In short, LLMs might be echoing common pitfalls of human systems engineering.

Over specified stakeholders primarily reflected solution seeking behavior that may be a common tendency for today’s LLMs [30]. Most of these referred to an organization or a facility that would not (or should not be) pre-determined during early-stage problem formulation. To elaborate, a large number of overspecifications were some aspect of the Mission Operations stakeholders, usually a specific facility or group, such as the Jet Propulsion Laboratory (JPL). However, they were still relevant to the problem statement and could be potentially viable stakeholders at a later stage in the ME lifecycle.

A significant error in the data set was the misidentification of NASA's Commercial Crew Program (CCP) as responsible for crew transportation to Ceres, a clear misunderstanding since CCP focuses solely on commercial partner access to the ISS. While CCP could theoretically be involved in such a mission, it is premature to label them as stakeholders, as mission specifics have not been defined.

Additionally, stakeholders like Transportation and Launch Operations, Collaborating Mission Assets, and Communication Networks were often overlooked, possibly due to the LLM's bias toward human stakeholders over infrastructure or environmental ones. Alternatively, these second order relationships may appear as distant to LLMs. Surprisingly, the US Government, the mission sponsor, was consistently missed, reflecting the LLM's inability to infer that NASA is funded by the government—a connection obvious to most but challenging for an LLM reliant on reference databases.

### 4.2. Variance in Responses

Table 2 shows that ChatGPT only finds a little over half of the ground-truth stakeholders correctly; however, it achieves this consistently. The implication being that the end user will receive some of the information they are seeking outright. This accompanied by a relatively low number of “wrong” stakeholders with a low standard deviation is indeed encouraging.

| Stakeholder Classification Across Parallel Threads | Mean | Standard Deviation |
|---|---|---|
| **Correct** | 6.76 | 1.56 |
| **Missed** | 5.24 | 1.56 |
| **Over Specified** | 4.80 | 2.72 |
| **Wrong** | 0.20 | 0.41 |
| **Total Stakeholders** | 11.8 | 3.19 |

*Table 2. Mean and Variation in Responses.*

The most concerning finding here is that the number of “missed” stakeholders was almost equal to the correct stakeholders and was equally consistent. This is coupled with a mean of 4.8 over specified stakeholders, close to the number of missed stakeholders; however, with a much larger standard deviation.

Combined with missed stakeholders, unusable information can represent one third to over half of the data in any given response. It is also plausible to expect that these responses are potentially underrepresented in this small data pool. Interestingly, the LLM did find an average of ~12 stakeholders, the same number in the ground truth; however, this was accompanied by a large standard deviation. This high variance seems to originate from seemingly random spikes in over specified responses that occurred intermittently. These findings imply that due to both abundance and variance of the missed, over specified, and wrong classifications, synthesizing correct responses will commonly be obfuscated in the general overall output.

## 5. Discussion and Conclusion

The complexity of modern SoS and challenges of planning for mission engineering activities that involve interrelated decisions under immense uncertainty necessitate domain-experts for effective problem formulation and solving. Emerging AI tools bring forth the promise alleviate some of expert workload and expedite these efforts, hopefully with increased quality and efficiency. However, translating open ended ill-defined problems into well-defined ones that are amenable for engineering development is an elusive challenge, and it remains to be seen how effective and consistent current LLMs are in assisting these goals. This paper presents preliminary results from a human-assisted stakeholder identification task for a NASA space mission problem to gauge how well and consistently a representative LLM can perform.

This study reports two main findings. First, in terms of quality, the glass is half full; as there is evidence to expect that multi-purpose AI tools such as ChatGPT could have some value in assisting problem formulation tasks [48]. We found that through iterative prompting, and purposefully no other further fine-tuning - a research choice at this stage - LLMs can provide a somewhat reasonable set of stakeholders. However, among this list, only half of the expected stakeholders are correctly identified, whereas the other half of expected stakeholders are consistently omitted. We observed that omitted stakeholders were usually external systems and operational/environmental factors. This was despite explicit prompting to guide the LLM to capture them. This may be a result of the lack of quality documents in training corpus. Arguably, it could be an extension of lack of distinction between the semantic meaning of what a "stakeholder means" thus indicative of how LLMs could exacerbate existing issues in the SE community as many SEs wrongfully only focus on humans and organizations when they study stakeholders. In addition, we also documented a strong tendency to over specify, often accompanied by an extensive list stakeholder that are remotely related to the mission. In some cases, but not always, this hinted at a solution seeking behavior, with stakeholders that would only be appropriate if a specific solution was to be selected. This is problematic given that we studied a problem formulation exercise and not a concept generation one.

The second finding is related to the ability of the LLM to provide *consistent* responses. Although we only experimented with a rather small data set of 25 independent threads, and a single LLM there was great variance in both the total number of stakeholders identified and the types of stakeholders captured in each independent attempt. LLMs are not deterministic tools, even with the expected stochasticity, we find this large variance concerning. This is more relevant given the tendency of practitioners to ask a given question once and take the given answers in verbatim, often without further verification or an analysis of variance. While in our case this could arguably be contributed to the uniqueness of the given mission statement - as it caused the LLM to do more "hunting" for an answer instead of pulling a well-vetted one as it would for a relatively simple task – there was never a repetition of outputs from the LLM nor was there mirroring in the phrasing of how stakeholders were described. Overall, these observations suggests that that every LLM response is "a roll of dice", and the quality of any given input could be subject to great variability; a key characteristic of LLMs that needs to be considered in future research.

Related to this concern, recent research notes that the mathematical reasoning of LLMs are potentially very fragile and may be the determining factor of high variance in repeated responses to the same inquiry [62]. While other disciplines are also starting to catch on to the issue of variance when working with LLMs, some are finding that as new LLMs emerge, some of this variance may be diminishing in certain applications [63], [64]. Regardless, this concern is not just intransient to mission engineering tasks and should be accounted for when evaluating any collaborative work done with current AI tools. Nevertheless, a much larger study would be required to understand how detrimental or possibly inconsequential this variance becomes for well-defined prompts for problem formulation exercises.

Finally, observing that the LLM is able to provide some good outcomes is encouraging, particularly for the kinds of problems ME is attempting to frame as these may not be available in a reference database, or could even be experienced for the first time by the given LLM. Problem uniqueness may be a major contributor to why the LLM is unable to provide a more complete answer. To that end, the contextually relevant knowledge of the practitioner to correlate new problems with existing processes is seemingly out of the grasp of LLMs currently. Nevertheless, ChatGPT does often provide a relatively useful set of responses that could be taken as an initial cut. Practitioners with an expertise in these skills could still benefit from the use of an AI assistant through the rapid generation of these tasks. This in return, could allow for a shift in cognitive load on the correction and refining of AI outputs rather than the pure generation of the task described. However, a non-expert would likely tend toward solution seeking behavior and potentially begin developing a system that satisfies the wrong stakeholders and needs.